\documentclass[conference]{IEEEtran}
\usepackage{textgreek}
\usepackage{amsmath,amssymb,amsfonts}
\usepackage{algorithm}
\usepackage{algorithmic}
\usepackage{makecell}
\usepackage{graphicx}
\usepackage{tabularx}

\usepackage{verbatim}
\usepackage{textcomp}
\usepackage{multirow}
\usepackage[table]{xcolor}
\def\BibTeX{{\rm B\kern-.05em{\sc i\kern-.025em b}\kern-.08em
    T\kern-.1667em\lower.7ex\hbox{E}\kern-.125emX}}
\begin{document}

\title{LinkBo: An Adaptive Single-Wire, Low-Latency, and Fault-Tolerant Communications Interface for Variable-Distance Chip-to-Chip Systems}

\author{\IEEEauthorblockN{Bochen Ye\IEEEauthorrefmark{1},\IEEEauthorrefmark{3}\thanks{\IEEEauthorrefmark{3}This work was performed while an intern at NXP.}, Gustavo Naspolini\IEEEauthorrefmark{2}, Kimmo Salo\IEEEauthorrefmark{2}, Manil Dev Gomony\IEEEauthorrefmark{1}}

\IEEEauthorblockA{\IEEEauthorrefmark{1}Eindhoven University of Technology, Eindhoven, The Netherlands}
\IEEEauthorblockA{\IEEEauthorrefmark{2}NXP Semiconductors, Nijmegen, The Netherlands}
bochen.ye@ed.ac.uk, \{gustavo.naspolini,kimmo.salo\}@nxp.com, m.gomony@tue.nl

 }
\linespread{0.85}
\maketitle

\begin{abstract}
Cost-effective embedded systems necessitate utilizing the single-wire communication protocol for inter-chip communication, thanks to its reduced pin count in comparison to the multi-wire I2C or SPI protocols. However, current single-wire protocols suffer from increased latency, restricted throughput, and lack of robustness. This paper presents LinkBo, an innovative single-wire protocol that offers reduced latency, enhanced throughput, and greater robustness with hardware-interrupt for variable-distance inter-chip communication. The LinkBo protocol-level guarantees that high-priority messages are delivered with an error detection feature in just 50.4 µs, surpassing current commercial options, 1-wire and UNI/O by at least 20X and 6.3X, respectively. In addition, we present the hardware architecture for this new protocol and its performance evaluation on a hardware platform consisting of two FPGAs. Our findings demonstrate that the protocol reliably supports wire lengths up to 15 meters with a data rate of 300 kbps, while reaching a maximum data rate of 7.5 Mbps over an 11 cm wire, providing reliable performance for varying inter-chip communication distances.

\end{abstract}

\begin{IEEEkeywords}
single-wire protocol, serial interface, hardware design, FPGA, Chip-to-chip communication.
\end{IEEEkeywords}

\section{Introduction}
\label{sec:intro}

\begin{figure}[t]
    \centering
    \includegraphics[width=1\linewidth]{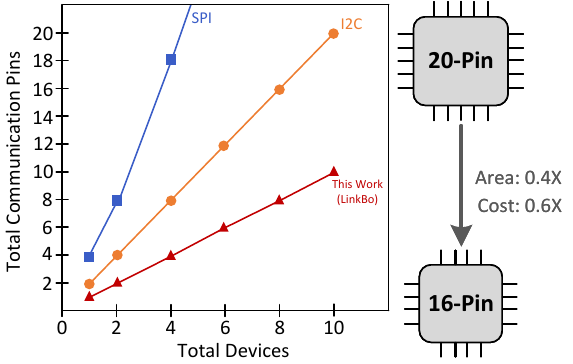}\\
     \makebox[0.48\textwidth]{\small (a) Figure A} 
     \\ \vspace{0.5em}
    \includegraphics[width=1\linewidth]{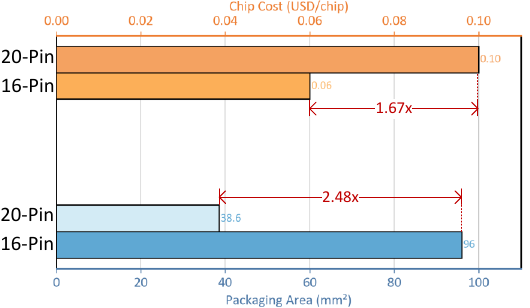}\\
    \makebox[0.48\textwidth]{\small (b) Figure B}
    \caption{The total communication pin-count cost with different communication protocol in multi devices system (left). The area and package cost between 20-pin package and 16-pin package (right). }
    \label{fig:em}
\end{figure}

Common  chip-to-chip communication methods typically rely on protocols such as Serial Peripheral Interface (SPI) or Inter-Integrated Circuit (I2C) \cite{spi}. SPI uses four wires: one for the clock signal, one for chip select, and two for data transmission. In contrast, I2C streamline communication with just two wires: one for data and one for the clock signal. However, as the number of interconnected chips increases, the total number of required pins for both SPI and I2C grows accordingly. In Fig.~\ref{fig:em} (left), the total number of communication pins required for SPI increases sharply while I2C exhibits a more linear growth.

In embedded systems, minimizing the physical footprint, particularly the number of off-chip communication wires, is a critical design goal. Reducing the number of external pins can significantly shrink the chip's packaging area, potentially saving several square centimeters. According to the packaging information~\cite{sot16,sot20}, a 16-pin package occupies only 40\% of the area of a 20-pin package and incurs just 60\% of the cost (Fig.~\ref{fig:em}). These findings highlight the strong correlation between pin count and both packaging area and cost. Consequently, reducing the number of external pins is essential for area-constrained applications. Although SPI and I²C require only 2 to 4 communication pins, they may still be suboptimal for systems with stringent area or cost limitations.

A single-wire communication protocol offers a promising alternative by single pin design. "Single-wire" refers to a class of protocol that uses a single wire for communication. The state-of-the-art (SOTA) single-wire protocol, known as the 1-wire protocol, operates as a half-duplex serial bus using only one data line and a single pin~\cite{adi}. However, the SOTA 1-wire protocol prioritizes long-distance communication, leading to extended bit-slot durations, which result in significant message latency and a low bit rate of 16.3 kbps, failing to meet the high-speed processing requirements of modern applications. Furthermore, 1-wire communication protocol is not particularly robust as it becomes susceptible to failure if the host malfunctions, and it lacks mechanisms to secure data integrity. To address these issues, this paper proposes the following contributions:

\begin{enumerate}
    \item We propose LinkBo protocol, which is a novel chip-to-chip single-wire protocol that can achieve lower latency, higher throughput, and improved robustness. It also contain a system-level model to analysis performance.
    \item We propose a novel hardware architecture for different speed and variable-distance LinkBo protocol on two FPGAs with only single wire. 
    \item Performance comparison of LinkBo with SOTA single-wire protocols in terms of latency and bit rate. In addition, the performance of the LinkBo protocol is analyzed under different parasitic parameters, wire lengths, and bit rate (clock frequencies) to assess robustness.
\end{enumerate}

The rest of the paper is organized as follows. 
Section \ref{sec:back} introduces the foundational concepts and examines related work on the single-wire protocol. 
Section \ref{sec:protocol} outlines the functionality of the Linkbo protocol. 
Section \ref{sec:model} offers a high-level overview of the channel model. 
Section \ref{sec:hardware} details the hardware architecture of our transmitter, receiver, and driver. 
Section \ref{sec:fpga} presents the hardware implementation on FPGA and verifies its correctness. 
Section \ref{sec:result} presents the FPGA setup and experimental results with respect to bit rate, latency, physical parameters, wire length, and clock frequency. 
Finally, the paper concludes with a summary of our findings and suggestions for future research in Section \ref{sec:concl}.

\section{Background and Related work}
\label{sec:back}
This section begins with an introduction to the current single-wire protocol, followed by a summary of related work.

\subsection{1-wire protocol overview} 
\begin{figure}[t]
    \centering
    \includegraphics[width=0.9\linewidth]{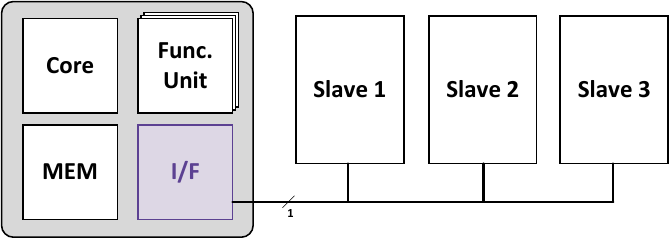}
    \caption{Current 1-wire bus architecture. The host contain processing core, function units, memory and 1-wire protocol. The host use 1-bit wire to send message to multiple slaves and receive data from slaves.}
    \label{fig:bus}
\end{figure}

The 1-wire bus~\cite{adi, ti}, which is a low-speed single-wire serial protocol designed for long-distance wired communication (extending up to 100 meters), is depicted in Fig.~\ref{fig:bus}. This bus architecture features a single host that exerts control over multiple slave devices. As illustrated in \cite{ti}, the communication process is initiated by the host transmitting a 960 µs reset pulse to all connected slaves. This is succeeded by a 64-bit device select message intended to address a specific slave, with each bit slot having a duration of 70 µs. Subsequently, the host may transmit either a 64-bit data message or a 16-bit command. In order to establish communication with an alternative slave, the host is required to recommence the reset pulse procedure. Communication between two chips is typically asynchronous due to different clock frequencies or unaligned clock edges, which can lead to errors at the receiver.To tackle this problem, it is essential to use either encoding techniques or clock synchronization strategies.

\subsection{Asynchronous communication}

The conventional 1-wire protocol uses low voltage to represent a 0 bit and high voltage to represent a 1 bit. This method introduces a DC component, which makes clock synchronization difficult and increases susceptibility to interference. In contrast, Manchester encoding eliminates the DC component, is self-clocking, and offers strong resistance to interference because each Manchester bit slot contains a transition. This transition can be utilized for both error detection and clock synchronization \cite{manchester}. 
\begin{figure}[htbp]
    \centering
    \includegraphics[width=1\linewidth]{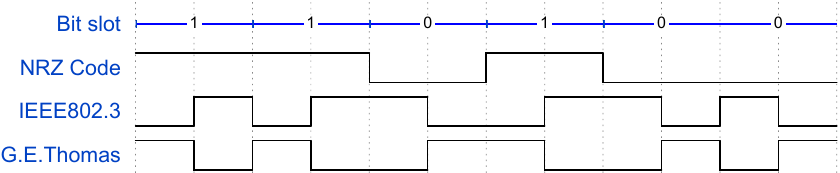}
    \caption{Different types of encoding method. Current 1-wire uses NRZ (Non-return-to-zero) code. IEEE802.3 standard of Manchester is just the opposite of the G.E.Thomas standard.}
    \vspace{-0.5cm}  
    \label{fig:manchester}
\end{figure}

\begin{figure}[t]
    \centering
    \includegraphics[width=1\linewidth]{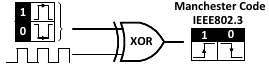}
    \caption{Encode method for IEEE802.3 standard Manchester code. The mask signal is periodic, while the data signal uses NRZ encoding.}
    \label{fig:encode}
\end{figure}

There are two standards for Manchester code: IEEE802.3 used in Ethernet and token bus network, the G.E.Thomas standard, which is the original version of Manchester encoding. Detail are shown in Fig.\ref{fig:manchester}. IEEE802.3 uses a rising edge to represent a 1 bit and a falling edge to represent a 0 bit, while the G.E.Thomas standard uses the opposite conventions. In this thesis, we use IEEE802.3 as our Manchester encoding standard, and the encoding method is illustrated in Fig.\ref{fig:encode}.
In Fig.\ref{fig:encode}, the mask signal is a periodic signal, and data signal is standard NRZ (Non-Return-to-Zero) code signal: high for a 1 bit, low for a 0 bit. Every Manchester bit must be transmitted within the period of the mask signal. Manchester encoding can be achieved using an XOR (exclusive OR) operation between the data and mask signals.

\subsection{Related work}
1-wire is a protocol first invented by Dallas Semiconductor (now part of ADI) in 1995 and is still used by TI and ADI today ~\cite{adi, ti}. 
Microchip developed a proprietary single-wire protocol known as the UNI/O bus \cite{unio}, which finds extensive application in embedded systems. Using Manchester encoding for synchronization, the UNI/O bus supports messages of arbitrary byte length and incorporates MAK/SAK acknowledgment. However, both bus systems experience inefficiencies due to extended standby pulses and an excessive number of address bits. Moreover, UNI/O does not include mechanisms for error checking or conflict resolution, deficiencies that our protocol addresses. There are also single-wire protocols designed for debugging systems rather than data transmission, such as SWIM \cite{swim} and debugWIRE. The comparison of the single-wire protocol is shown in Table.\ref{tab:related}. 

Dos Reis et al. \cite{DCU} devised a one-wire vehicle network with a ring topology capable of handling multi-byte messages and acknowledgment bits; however, it operates at a modest speed of 38.4 kbps. Rahman et al. \cite{multiplex} merged data and clock signals on a single wire in hardware, but only achieved enhancements over the minimum bit rate. Rezaei et al. \cite{wfm} employed FDMA for simultaneous data transmission, yet this led to increased errors as the number of nodes increased. Nikolic et al. \cite{cdma1} implemented CDMA in a VLSI bus but encountered difficulties adapting it to a single-wire configuration. Furthermore, studies \cite{cdma}, \cite{cdma2} explored one-wire CDMA-based buses; the first experienced high error rates with node counts greater than seven, while the second only offered a solution simulated through software.
Most papers use the single-wire protocol for embedded applications, such as sniffer devices \cite{sniff}, water quality monitor \cite{fpga} and slave interface \cite{unv}. Others analyze and evaluate the single-wire protocol using various methods \cite{analysis}. However, these works do not improve the protocol, only utilize it.
Some paper optimize single-wire interface in analog circuit while our work only use digital technology~\cite{FDMIF,FDMIF2,srrt}.
   
\begin{table}[htbp]
    \caption{Comparison of different single-wire protocols.}
    \begin{center}
    \begin{tabular}{|c|c|c|c|}
    \hline
    Protocol&\makecell{1-wire\\(TI,ADI)}&\makecell{UNI/O\\(Microchip)}&\makecell{SWIM/\\debugWIRE} \\
    \hline
    Sync/Async&Async&Async&Async\\
    \hline
    Speed&8.33-111kbps&10-100kbps&10lbps-1Mbps\\
    \hline
    Type&Master/slave&Master/slave&Point to Point\\
    \hline
    Duplex&Half&Half&Half\\
    \hline
    Pin count&1&1&1\\
    \hline
    Application&EEPROM,sensor&EEPROM,sensor&Debug system\\
    \hline
    \end{tabular}
    \label{tab:related}
    \end{center}
\end{table}
  
\section{LinkBo protocol definition}
\label{sec:protocol}
In this section, we provide an overview of our protocol and then introduce its main features.
\subsection{Protocol overview}

\begin{figure*}
    \centering
    \includegraphics[width=1\linewidth]{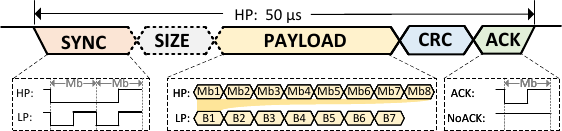}
    \caption{Two priority LinkBo messages format. HP and LP messages have different SYNC and PAYLOAD fields, but share the same CRC and ACK fields. Only LP messages include a SIZE field. HP messages are always completed within 50~µs, whereas the duration of LP messages depends on the payload size. In PAYLOAD, one byte in LP message contain 8 Manchester bits.}
    \label{fig:message}
\end{figure*}

The proposed LinkBo protocol uses Manchester encoding (IEEE802.3) to enable bidirectional data transmission between two asynchronous chips over a single wire. Fig.~\ref{fig:message} illustrates the message format and the specific timing details of the LinkBo protocol. Each message is composed of multiple fields: synchronization field, size field, payload field, CRC field, acknowledgment field. The detail:
\begin{enumerate}
    \item Synchronization field: This initial field synchronizes the receiver with various clock signals, and it also serves for interrupt purposes, with further explanation provided later. 
    \item Size field: The size field is optional and varies depending on the message, indicates the number of bytes to be sent in a single message. 
    \item Payload Field: The payload field carries the actual data information. 
    \item CRC field: This field adds four CRC bits at the end of the message, which is used by the receiver to check for errors. 
    \item Acknowledgment field: The acknowledgment (ACK) field provides a response from the receiver to the sender, indicating whether the transmission was successfully completed.
\end{enumerate}

\subsection{Priority message and interrupt}
In the LinkBo protocol, messages are categorized into two types based on the structure of their synchronization fields, as illustrated in Fig.~\ref{fig:message}.
\begin{enumerate}
    \item The first type is the high-priority (HP) message, which uses only 15 Manchester bits (Mb): 2 bit slots for synchronization, 8 bits for the payload, 4 bits for CRC, and 1 bit for acknowledgment (ACK). Each HP message can carry only one byte (B) of payload and does not include a size field. 
    \item The second type is the low-priority (LP) message, which can support 1 to 7 bytes of payload. It uses a 3-bit size field to represent this. Consequently, a LP message contains from 18 to 66 Mb: 2 bits for synchronization, 3 bits for size, 8 to 56 bits for the payload, 4 bits for CRC, and 1 bits for ACK.
\end{enumerate}

HP messages are designed to support hardware interrupts, providing minimal latency for time-critical or emergency data transmission but carries only a small amount of information. To balance throughput, LP messages are introduced to transmit larger volumes of data, albeit with higher latency. 
LinkBo allows HP messages to interrupt LP messages transmission when necessary to ensure low latency. This hardware-level interrupt enables faster response times compared to software-based interrupt handling (e.g., I2C), ensuring timely delivery of high-priority data.

\subsection{Synchronization and re-synchronization}

In single-wire communication, synchronization must rely solely on signal edges due to the absence of a dedicated clock line. To differentiate between HP and LP messages during synchronization, the LinkBo protocol utilizes two distinct synchronization field patterns.
As illustrated in Fig.~\ref{fig:message}, the synchronization field includes two Mb slots. For HP messages, the initial slot is consistently held low to differentiate it from LP messages. This extended low signal also facilitates the creation of an interrupt signal on the bus, enhancing its detectability relative to other signals. In contrast, for LP messages, the first slot signifies a 1, marked by a transition from low to high. The second Mb in both types of message is the same and represents a 1.

Upon detecting the first falling edge, the prescaler counter (PSC) starts to measure the cycle duration. If the low signal lasts longer than one Manchester slot, the message is identified as HP; otherwise, it's LP. For HP messages, the counter stops at the first rising edge; for LP messages, at the second. The count reflects 1.5 Mb slots, so dividing by 1.5 yields the Mb slot duration in the receiver’s clock domain. This process is known as synchronization.
\begin{figure*}
    \centering
    \includegraphics[width=1\linewidth]{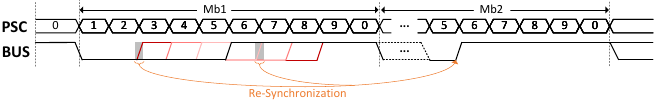}
    \caption{The re-synchronization process for Manchester code during the transmission. The depth of the red edge represents the severity of the error.}
    \label{fig:resync}
\end{figure*}
    
Manchester coding requires a transition in the middle of every Mb slot, which can be used for error checking. In Fig.~\ref{fig:resync}, black signal shows a correct Manchester code with a transition in the middle, indicating that there are no errors. The red signal shows early or late transitions, which are not correct but still acceptable. If edge in other place, it represents a error case for the receiver, where it cannot determine whether the Manchester bit is 1 or 0. Resynchronization is the process of adjusting the next transitions to ensure that they occur in the middle of each Mb slot. This mechanism ensures that each Manchester bit is correctly aligned, helping to maintain accurate data transmission.

\subsection{CRC and acknowledgment}

Cyclic redundancy check (CRC) is a method used to detect errors in data messages or memory storage. By performing a polynomial division on the data to generate an N-bit checksum, the CRC can effectively detect and identify common errors such as bit flips, ensuring the integrity and reliability of the data \cite{crc}. In this work, we use a 4-bit CRC with the polynomial $X^4+X+1$ because it can detect any single-bit error and can detect most two-bit error. In Eq.~\eqref{for:crc}, the $b$ represents the number of burst error bits, and $k$ means the number of CRC bits. It shows that a 4-bit CRC can detect burst errors of up to 4 bits with 100\% accuracy and burst errors longer than 4 bits with 93.75\% accuracy.
\begin{equation}
\label{for:crc}
    P_{crc}(b)=
    \begin{cases}
    1, &\text{if } b \leq k\\
    1-\frac{1}{2^k},&\text{if } b >k
    \end{cases}
\end{equation}
    
The receiver will then send an ACK signal back to the transmitter. If the CRC check is correct, it will send Manchester code 1 (low to high). If the CRC check is incorrect, it will not edge. The ACK signal is shown in Fig.~\ref{fig:message} right.

\subsection{Scalability for multi-device}
\begin{figure}
    \centering
    \includegraphics[width=1\linewidth]{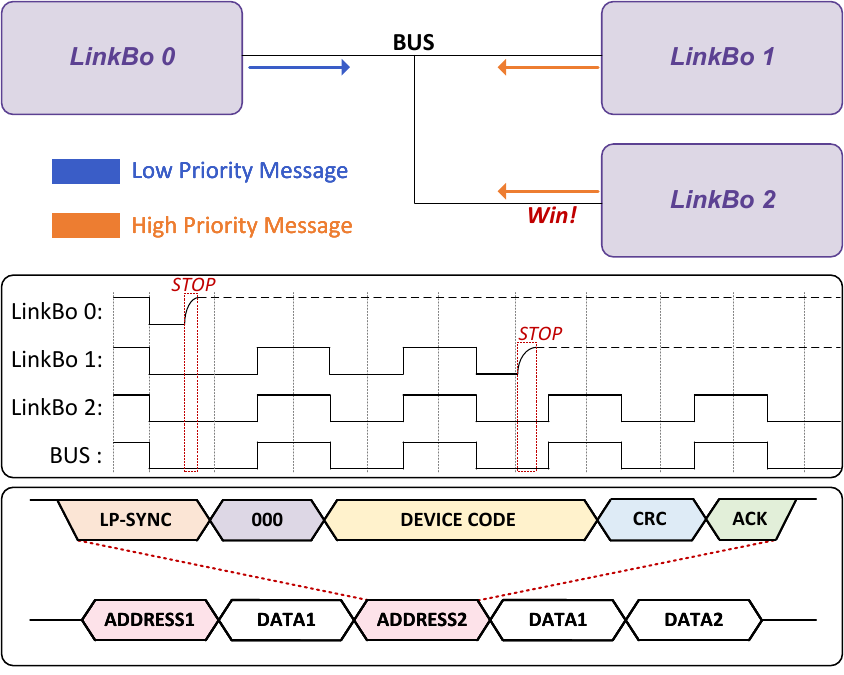}
    \caption{(a) Multi-device arbitration with Wired-AND logic. (b) The address message for configuration.} 
    \label{fig:multi}
\end{figure}
The 1-Wire bus operates with a single master and multiple slaves, which limits scalability and places a heavier load on the host. If the host device fails, the entire system stops working. To address this, the LinkBo protocol uses a peer-to-peer bus, ensuring better scalability and fault tolerance. In LinkBo, each module can communicate independently with others, allowing the system to continue functioning even if one device fails. 
In Fig.~\ref{fig:multi} (a), The bus uses wired-AND arbitration during transmission. LinkBo 0 sends a low-priority message, while LinkBo 1 and 2 send high-priority ones. From the bus's perspective, the first high signal loses arbitration, so LinkBo 1 and 2 are excluded successively. To support address in our protocol, we can repurpose the reserved ‘000’ SIZE field to define a new message type that carries an address in Fig.~\ref{fig:multi} (b). Each device responds only to its unique address. This approach adds latency but minimizes hardware changes to the original design.

\section{System model for performance evaluation}
\label{sec:model}
This section presents the high-level baseline model of the 1-wire protocol and the ideal transmission model of the LinkBo protocol, implemented in Simulink. It then discusses incorporating a channel model into the ideal model to better approximate real-world conditions.
\subsection{Baseline model}
To start with the baseline model, current 1-wire (ADI) protocol is selected, as it is widely used and easy to implement. The operation of this protocol has already been introduced in Section \ref{sec:back}. 
    
The first synchronization message in the 1-wire protocol consists of an 8-bit family code, a 48-bit address code, and an 8-bit CRC code. The family code specifies the device family, theoretically allowing the 1-wire bus to connect up to $2^{56}$ devices. After the synchronization message, a specific device is selected to receive commands from the host. These commands are typically 16-bit ROM commands or functional commands. Devices that are not selected will ignore the message until the next reset pulse, at which point they can be selected again.

After analyzing the baseline model, we identified several drawbacks of the 1-wire bus. To address these issues, we developed solutions and improvements which have been incorporated into our LinkBo protocol.
    
One of the significant drawbacks of the 1-wire protocol is its long message duration. In our Simulink model, the reset message takes 960 µs, and each write and read bit slot requires 60 µs (no 10 µs recovery time). The synchronization message takes 4.8 $ms$. This long-time message causes the latency and low bit rate. For the baseline model, the bit rate is 16.26 kbps initially, but it stabilizes at 16.66 kbps for subsequent operations. To address this issue, we need to reduce the length of the synchronization message and shorten the bit slot time. While the CRC field is useful for error detection, an 8-bit CRC may be excessive, so reducing the CRC field length is also necessary.

Another drawback is that the host cannot immediately verify if the device has received the message correctly. The host only receives data returned by the device. If an error occurs during transmission and the device does not write data to the bus due to not receiving the correct command, the host will continue sending erroneous data. Although this represents a worst-case scenario, the protocol needs an acknowledgment mechanism to improve robustness and ensure reliable communication.

The last drawback is that only one host can control the bus at a time, preventing devices from communicating directly with each other. In the baseline model, adding more slave devices increases the management burden on the master device, as all data must be routed through it. This can lead to the master becoming a performance bottleneck, especially in applications involving large amounts of data, which places a heavy load on the host. To address this issue, we propose a peer-to-peer approach where each device can actively initiate communication. This change makes the network more flexible and reduces dependency on a single master device, improving overall efficiency and performance.
    
\subsection{Ideal LinkBo model}
\begin{figure}
    \centering
    \includegraphics[width=1\linewidth]{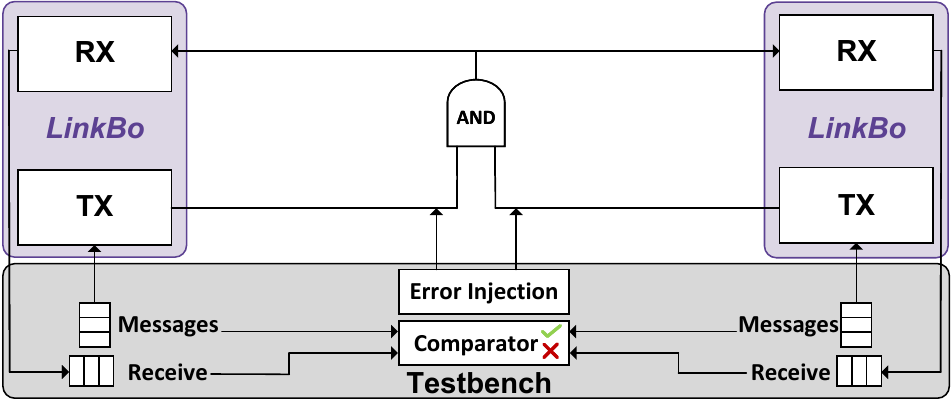}
    \caption{Ideal LinkBo model and testbench. The testbench can send messages to any module and receive them from another module.}
    \vspace{-0.5cm}  
    \label{fig:ideal}
\end{figure}

After exploring and analyzing the baseline model, we implemented an ideal LinkBo model in Simulink, as shown in Fig.\ref{fig:ideal}. The model includes two identical LinkBo modules, each comprising a transmitter (TX), a receiver (RX), and a driver. Both the TX and RX use the driver to output digital signals that control the 1-wire bus. 
    
To simulate a real 1-wire bus, we use an AND gate to model the wired-AND logic on the bus. For instance, when two blocks attempt to send data simultaneously, the first device that pulls the bus to a low voltage against the other devices connected to the bus will dominate the bus, and this signal will be received by the other module's RX. This is used solely for the behavior of LinkBo, and a detailed channel will be presented in the next section.

To test our model, we constructed a testbench for the ideal LinkBo model. This testbench is designed to provide stimuli to the TX of one LinkBo module and receive messages from the RX of another. Additionally, to evaluate functional behavior, the testbench can inject errors or noise into the bus. 

\subsection{Channel model}
\begin{figure}
    \centering
    \includegraphics[width=1\linewidth]{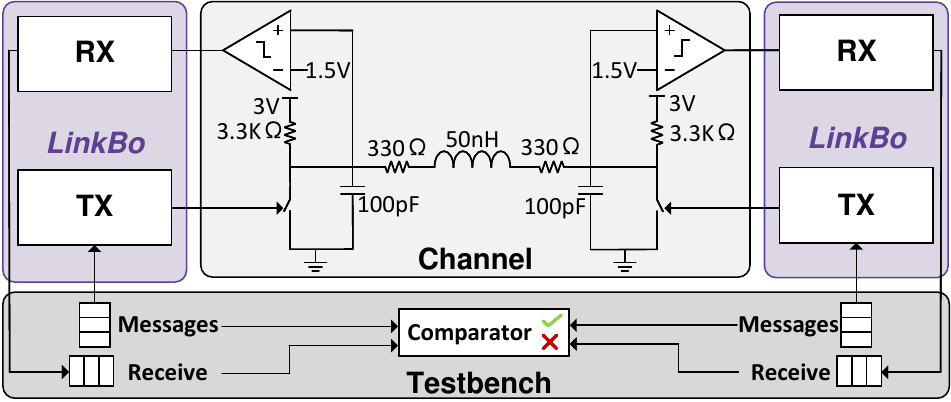}
    \caption{Channel model with parasitic parameter. In this figure, the TX and RX serve as interface between LinkBo and channel. The inductance, resistance and capacitance simulate the real wire characteristics.}
    \label{fig:channel}
\end{figure}

Hardware inherently possesses numerous parasitic parameters that can influence voltage and current delivery, resulting in performance variations in actual chips. To investigate the performance of the LinkBo protocol, we developed a high-level model in Simulink, depicted in Fig.~\ref{fig:channel}. This model features two identical LinkBo modules, each with a transmitter (TX) and a receiver (RX). Both the TX and RX leverage the driver to transmit digital signals to control the single-wire bus. For testing our model, we set up a test bench for the ideal LinkBo model, designed to deliver stimuli to the TX of one module and receive data from the RX of the other.

In Fig.~\ref{fig:channel}, the TX is connected to a switch, while the RX is connected to a 1.5V comparator. When the driver is high (bit 1), it opens the switch, allowing current to flow through the channel. When the digital output is low (bit 0), the switch closes, directing the current to the ground. If the voltage exceeds 1.5V, the input of RX registers as binary level 1; otherwise, it registers as binary level 0.

\begin{figure}[t]
    \centering
    \includegraphics[width=1\linewidth]{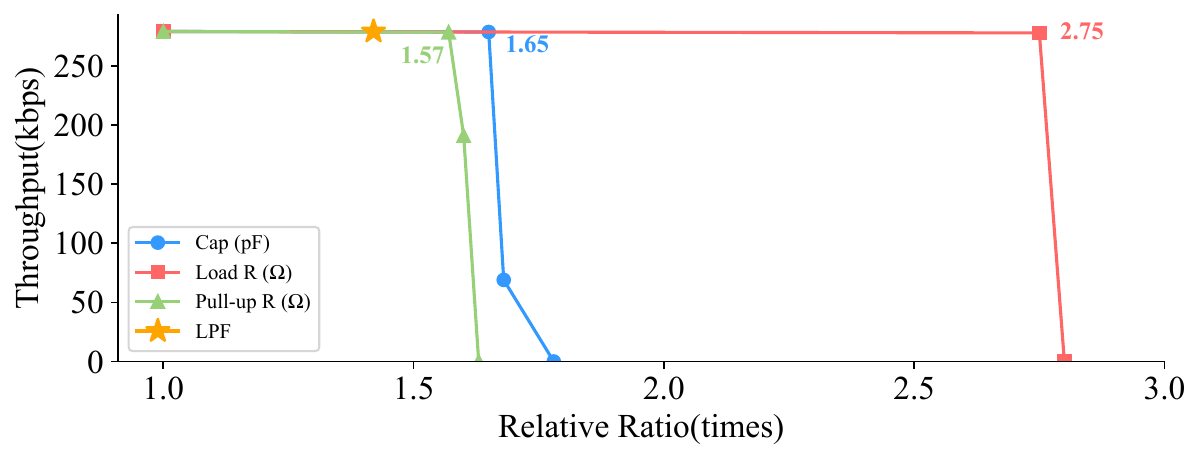}
    \caption{Throughput vs Parameter (Capacitance, Load resistor, Pull-up resistor). All three lines drop sharply at a certain point, but they can still reach the LPF condition without any performance degradation.} 
    \label{fig:4f}
\end{figure}

To ensure that our protocol can achieve the same throughput or bit rate in a real chip, the parasitic parameters of the channel model can be adjusted to evaluate the relationship between parasitic parameters and throughput. These parameters are the same as those used on the PCB that holds the FPGA.
Fig.~\ref{fig:4f} presents the simulation results, with the X-axis indicating the parameter variation ratio. On the Y-axis, $Throughput$ is depicted, defined as the count of RX bits received successfully per unit time. In Eq.~\eqref{for:3}, $B_{re}$ denotes the bits received by RX, and $T$ represents the duration of one transmission.
\begin{equation}
\label{for:3}
    Throughput=\frac{B_{re}}{T}
\end{equation}  

The blue capacitance line signifies a tolerance for a 1.65X variation; however, its efficiency drops to 25\% at a 1.68X variation before gradually declining to zero. The red line for the load resistance shows that it can withstand a 2.75X variation before rapidly falling. This is because a change in the load resistor alters the current while the capacitor impacts the circuit's rise and fall times, delaying voltage stabilization. The green pull-up resistor line demonstrates a sharp performance decline following a variation of 1.57X. Due to the size of the baseline pull-up resistor, slight changes significantly affect resistance. Inductance, not depicted in Fig.~\ref{fig:4f}, can support up to 167.5 µH. In summary, the component that affects the throughput is mainly the capacitor, followed by the pull-up resistor, then the load resistor, whereas the inductance has the least influence.

The maximum resistance and capacitance for the low-pass filter to filter out high-frequency noise can be calculated by Eq.~\eqref{for:cutoff}. $F_c$ is cut off frequency set at 3.3 MHz. As a result, the load resistor needs 470 $\Omega$, while the capacitance needs 100 pF. The 470 $\Omega$ is only 1.42X of the baseline value (star marker in Fig.~\ref{fig:4f}), so it can easily meet the requirement.
\begin{equation}
\label{for:cutoff}
    F_c=\frac{1}{2\pi RC}
\end{equation}

\section{Hardware architecture and implementation}
\label{sec:hardware}
\subsection{Top-level architecture}

\begin{figure*}[t] 
    \centering
    \includegraphics[width=1 \linewidth]{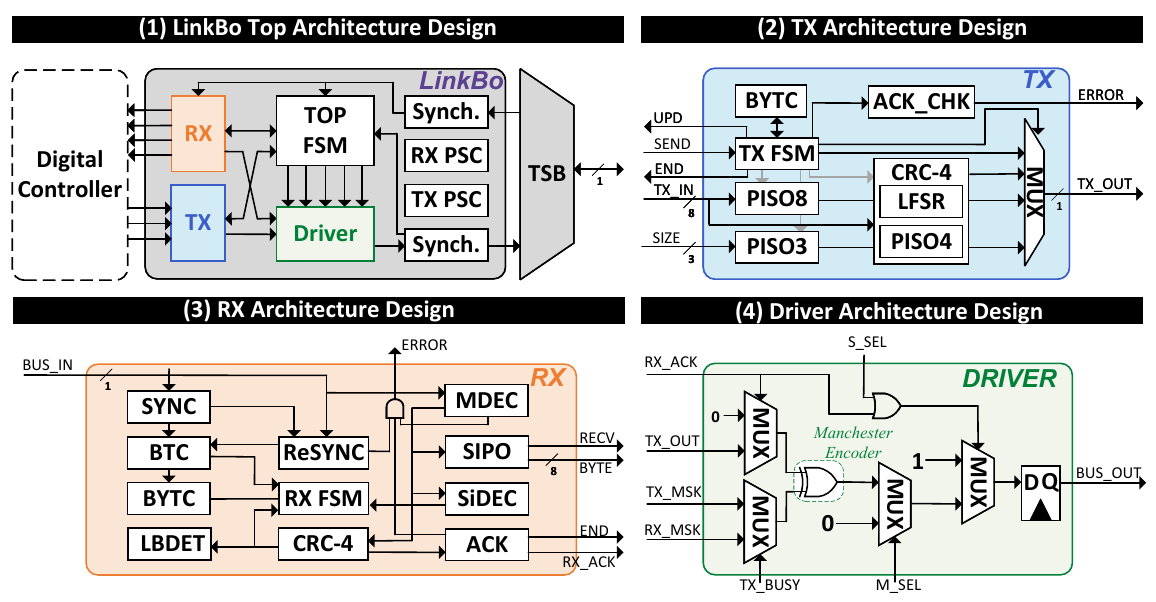}
    \caption{(1) Top-level architecture. LinkBo module is part of digital system in each chip, consisting of synchronizer, TOP FSM, RX, TX, driver, and PSC(for both TX and RX). (2) Transmitter architecture. TX is responsible for sending any priority message. (3) Receiver architecture. The RX is responsible for receiving messages, synchronizing, and sending an acknowledgment (ACK) signal. (4) Driver architecture. Driver consists of one Manchester encoder (XOR gate) and 4 MUXes.} 
    \label{fig:all}
\end{figure*}

In Fig.~\ref{fig:all} (1), the top-level hardware architecture of LinkBo is shown. The core elements within the LinkBo module include the transmitter (TX), receiver (RX), prescaler counter (PSC), TOP FSM, synchronizer, and driver. The TX module sends messages, while the RX module is in charge of receiving them. The driver encodes information into Manchester bits and controls the bus. The PSC divides the clock frequency and produces the Manchester mask signal. The synchronizer resolves timing discrepancies in digital circuits when signals transition between distinct clock domains. The TOP FSM, a state machine, manages the activation and deactivation of TX/RX operations during interruptions. There is a tri-state buffer out of LinkBo module, which is used to control the input and output of bus. The digital controller can provide the LinkBo module with different clock frequencies to support various speeds or bit rates for communication over different distances.

\subsection{Transmitter architecture}

The transmitter is controlled by TX FSM. The UPD signal confirms that the $BUS\_IN$ has been transmitted. Two shift registers, an 8-bit and a 3-bit, convert parallel input data into 1-bit serial output data (PISO8, PISO3). The CRC-4 module employs 4 registers in a Linear-Feedback-Shift-Register (LFSR) configuration with the polynomial $X^4+X+1$ to produce a 4-bit CRC checksum, transmitted bitwise through a 4-bit shift register (PISO4). A 4-to-1 multiplexer (MUX) selects the data set for transmission. The byte counter (BYTC) can ensure to send multi-byte message. The $ACK\_CHK$ module verifies CRC errors and reports errors if found. The architecture is depicted in Fig.~\ref{fig:all} (2).
      
\subsection{Receiver architecture}

The receiver is depicted in Fig.~\ref{fig:all} (3) and is more sophisticated than the transmitter. The synchronization module (SYNC) first identifies the falling edges, initiating a counter to determine the Manchester bit slot; as previously explained in Section \ref{sec:protocol}. The Manchester decoder (MDEC) converts each Manchester bit into a standard bit, while the serial-in-parallel-out (SIPO) module delivers one byte of the message. The CRC-4 divider checks for errors in the payload. If none is found, the ACK module signals the driver. The state machine (RX FSM) utilizes bit and byte counters (BTC, BYTC) to track Manchester bits and message bytes. Sometimes, edges may not align centrally; the re-synchronization module (Re-SYNC) is designed to address this. Lastly, the low-bus detector (LBDET) watches for prolonged low-level signals on the bus, signaling a high-priority event and triggering an interrupt.
      
\subsection{Driver architecture}

The driver consists of several MUXes and a Manchester encoder, see in Fig.~\ref{fig:all} (4). The Manchester encoder uses an XOR gate to generate Manchester code. The driver utilizes an output register to eliminate race conditions and hazards in the combinational logic.

\section{FPGA Implementation and verification}
\label{sec:fpga}
\begin{figure}[t]
    \centering
    \includegraphics[width=1\linewidth]{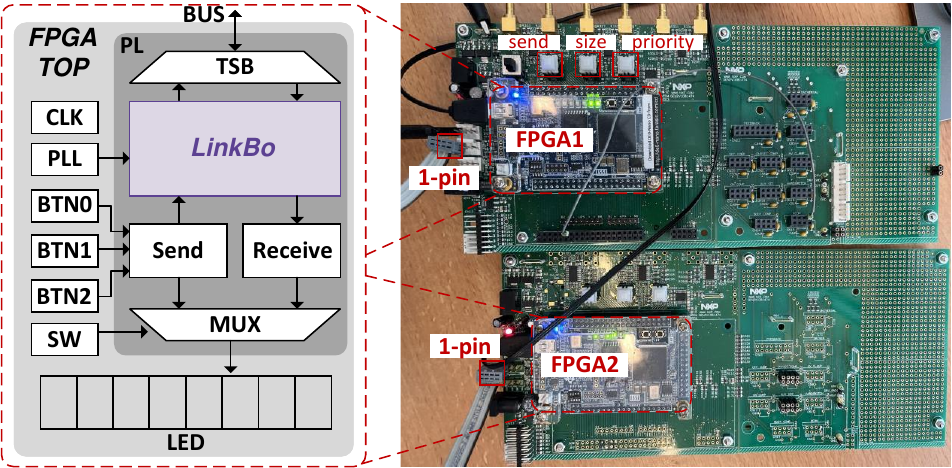}
    \caption{FPGA test setup. Two FPGAs are connected via a single wire, sharing the same PL design. Buttons and switches are used for signal input, while LEDs display the transmitted and received data.}
    \label{fig:fpga}
\end{figure}
In this section, the LinkBo hardware design implemented on two FPGAs with single wire and verify its communication signal.

Fig.~\ref{fig:fpga} shows the FPGA test setup. In our experimental setup, two Altera Cyclone-IV FPGAs are used and connected via a single pin wire. Each FPGA features three buttons (BTN) to pick input signals (send, size, and priority) and eight LEDs to indicate payload data. FPGA 1 transmits data to the LinkBo module, showing the size and priority of the LEDs, while FPGA 2 receives data from the LinkBo module, with the LEDs showing a payload byte. A Phase-Locked Loop (PLL) can control the clock frequency up to 100MHz and give it to LinkBo module (3MHz in this test). Tri-state buffer (TSB) manages the bus output or input on the shared communication line, and a pull-up resistor of PCB ensures that the bus defaults to a high voltage when unconnected.

\begin{figure}[t] 
    \centering
    \includegraphics[width=0.5\textwidth]{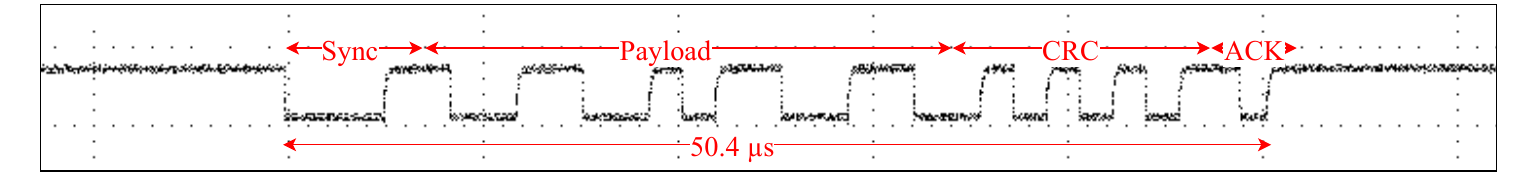}
    \\
    \makebox[0.235\textwidth]{\small (a)}
    \\
    \includegraphics[width=0.5\textwidth]{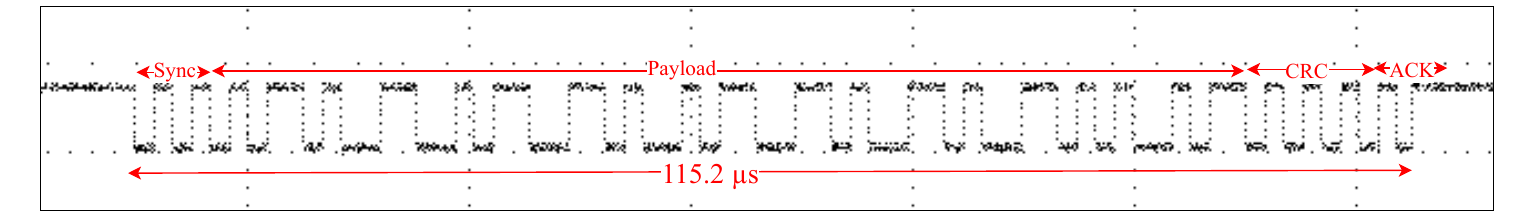}
    \\
    \makebox[0.235\textwidth]{\small (b) }
    \\ 
    \includegraphics[width=0.5\textwidth]{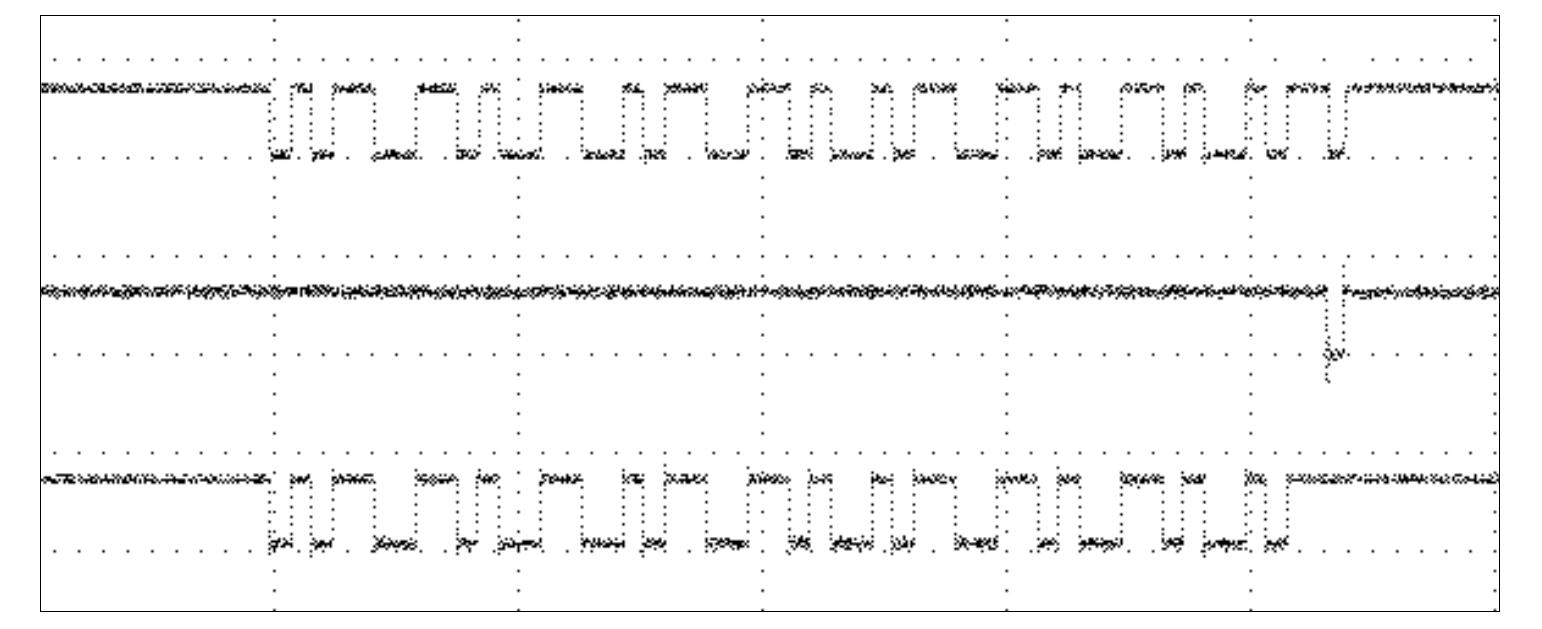}
    \\
    \makebox[0.235\textwidth]{\small (c) }
    \caption{Different waveform signals in oscilloscope. (a) High-priority message signal in oscilloscope. (b) Low-priority message signal in oscilloscope. (c) Three signals (BUS,ACK, TX output) in oscilloscope.} 
    \label{fig:3f}
\end{figure}

In FPGA implementation, we can use an oscilloscope to detect real waveform on the 11 centimeter wire. In Fig.~\ref{fig:3f} (a), the HP message signal sent from FPGA 1 to FPGA 2. The HP message takes 47.5 µs and ACK use 2.9 µs. In Fig.~\ref{fig:3f} (b), LP message carrying a 3-byte payload takes 112.4 µs and ACK use 2.8 µs. In Fig.~\ref{fig:3f} (c), there are three different signals are shown: the top one is bus signal, the middle one is bus out from FPGA 2 at 2.85 MHz, the last one is bus out from FPGA 1 at 3.15 MHz. The result shows our protocol and hardware can handle the asynchronous communication at $3MHz \pm 5\%$. The FPGA 2 will send back ACK signal only when it check CRC is right, which introduces a slight delay.

\section{Experiments and results}
\label{sec:result}
In this section, we performed several comparisons and experiments using FPGA hardware. The purpose of these experiments is to illustrate low-latency and robustness of LinkBo protocol.

\subsection{Latency analysis}

\begin{figure*}
    \centering
    \includegraphics[width=0.8\linewidth]{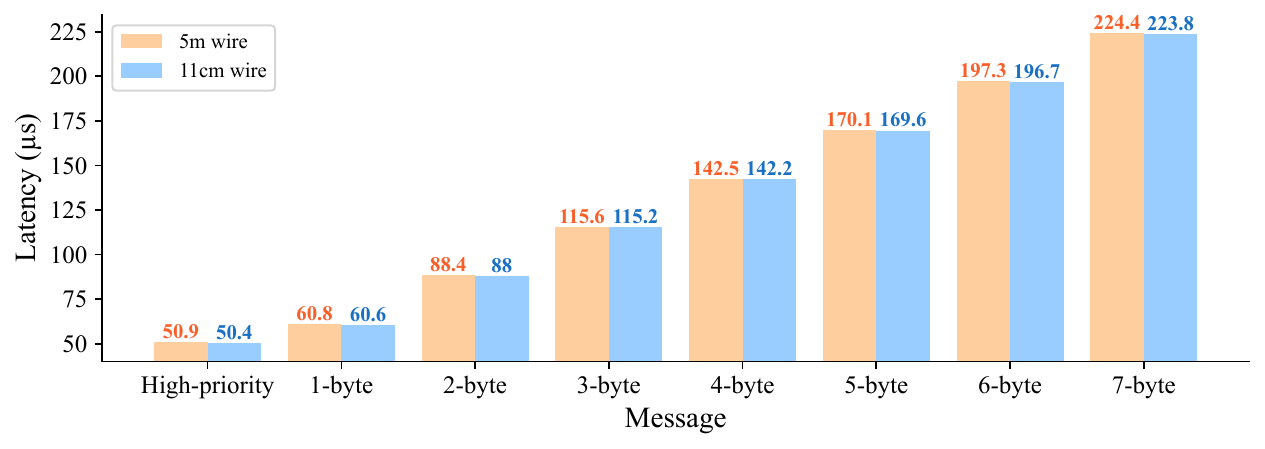}
    \caption{Latency for different messages. The HP message is transmitted faster than the LP message, even though they carry the same payload. The wire length does not significantly affect latency.}
    \label{fig:latency}
\end{figure*}

Fig.~\ref{fig:latency} shows that sending HP messages yields a minimum latency of 50.4 µs, while LP ones take 60.6 µs for the same bit number. In addition, LP message latency increases linearly with payload size, peaking at 224.4 µs for a 7-byte message. Comparison of wire lengths shows that they have a minimal effect on latency.

\subsection{Sensitivity analysis}

\begin{figure}[t]
    \centering
    \includegraphics[width=1\linewidth]{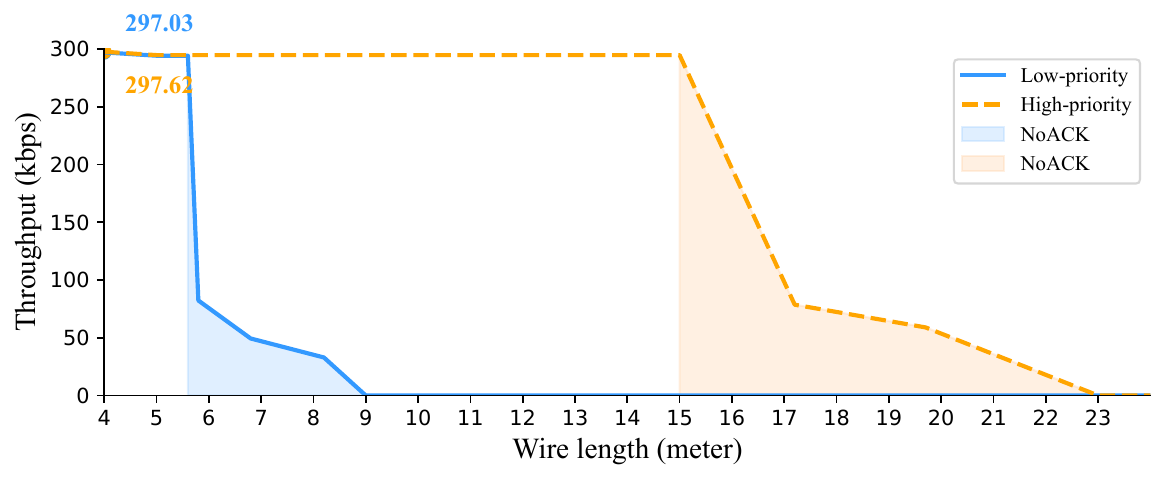}
    \caption{Throughput of 2 priority message vs wire length. The throughput remains stable up to a certain threshold length, after which it suddenly drops to a low level and then gradually approaches zero. The shadow means RX not give back ACK signal.}
    \label{fig:wirelength}
\end{figure}

To assess performance, the length of the cable between two FPGAs is varied, starting with a 5 meter wire for data transfer. Fig.~\ref{fig:wirelength} reveals that LP message performance declines rapidly between 5.6 and 9 meters, eventually reaching zero. However, HP messages begin to decline at 15 meters, reaching zero at 23 meters. This discrepancy arises because the synchronization pulse for HP messages is three times longer than that for LP ones, increasing robustness. This also accounts for the 1-wire (ADI) protocol's requirement for longer reset times and bit slots. When the message length exceeds a threshold, signal degradation prevents the receiver from reconstructing a full byte, thereby no ACK on the bus. The curve in Fig.~\ref{fig:wirelength} is similar to that in Fig.~\ref{fig:4f}, showcasing a sharp drop beyond a certain point. Unlike the previous high-level test that altered only one parameter, varying the length of the wire in this experiment can affect the inductance, the load resistor, and the load capacitance concurrently.

\begin{figure}[t]
    \centering
    \includegraphics[width=1\linewidth]{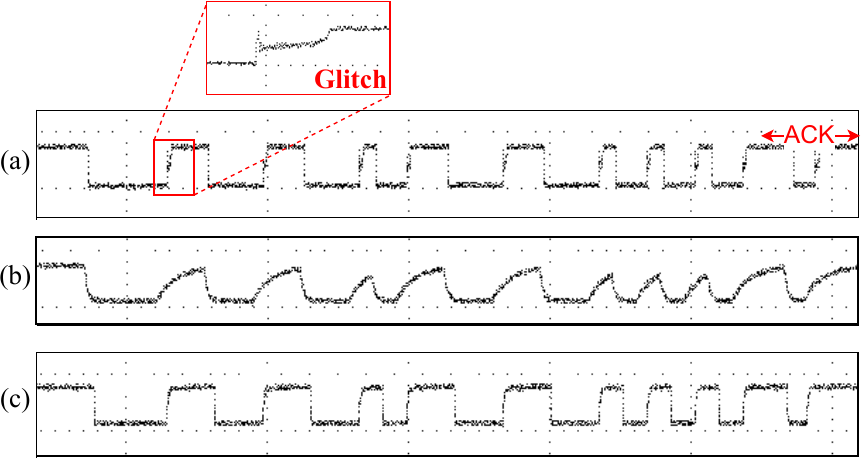}
    \caption{The waveform signal of HP message in different wire length by oscilloscope. (a) Waveform of output pin with 14.8 meter. (b) Waveform of 14.8 meter bus wire. (c) waveform of 11 centimeter bus wire.}
    \label{fig:wave3}
\end{figure}
In Fig.~\ref{fig:wave3}, (a) presents the output pin waveform when two FPGAs are linked by a 14.8-meter cable, highlighting glitches on the rising edge. (b) shows the waveform of a bus signal over 14.8 meters, while (c) illustrates it over an 11-centimeter distance. Findings show that longer wires shorten high-level duration, extend rise time, and eventually hinder edge detection beyond a threshold length.

\begin{figure}[t]
    \centering
    \includegraphics[width=0.7\linewidth]{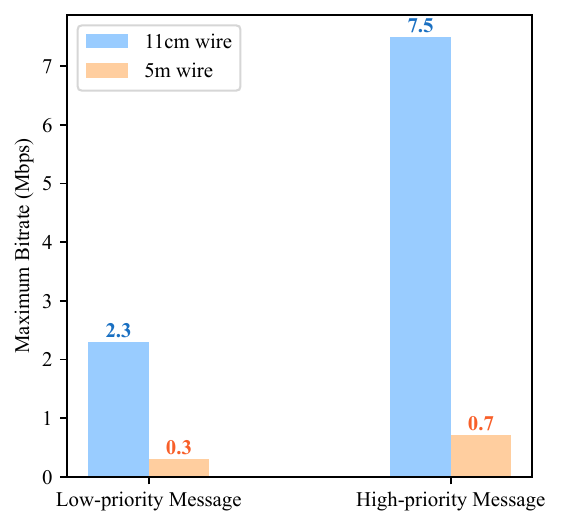}
    \caption{Maximum bit rate of different priority message for 11cm and 5m wire length. The longer the wire length, the lower the supported bit rate. High-priority messages can support a bit rate that is 2 to 3 times higher than that of low-priority messages.}
    \label{fig:maxfreq}
\end{figure}

Our experiments demonstrate that LinkBo supports distances up to 5.6 meters for LP messages and 15 meters for HP messages, which is noticeably shorter than what 1-Wire (ADI) offers. However, LinkBo is optimized for chip-to-chip communication on a single PCB or multiple PCB rather than for extended distances. As chip-to-chip communication generally occurs over centimeters, LinkBo reduces bit slot duration and elevates clock frequency, thus achieving lower latency and a higher bit rate. As detailed in fig.~\ref{fig:maxfreq}, the highest bit rate reaches about 300 kbps (3 MHz) for low-priority messages and 710 kbps (7.1 MHz) for high-priority messages over a 5-meter wire, although such lengths are impractical for PCBs. At an 11 cm scale, low-priority messages can attain up to 2.3 Mbps, while high-priority messages can reach 7.5 Mbps. This indicates that our protocol provides very high bit rates over short distances and maintains strong performance over 10–20 meters.

\subsection{Protocol comparison}

Table ~\ref{tab:compa} shows a comparison of this work with the SOTA protocol in terms of latency and robustness. Half-duplex refers to devices that transmit in one direction at a time. Bit rate specifies the bits sent per second. In Eq.~\eqref{for:1}, Bit rate $R_b$ represents the kilobits per second, $B_{total}$ is the total bits per transmission, and $T$ is the total duration of a transmission.
\begin{equation}
\label{for:1}
    R_b=\frac{B_{total}}{T}
\end{equation}
The effective bit rate (EBR) denotes the upper limit for payload data bits transferred per second. In Eq.~\eqref{for:2}, $B_{data}$ signifies the total number of data bits within a single transmission. Latency is defined as the minimum duration for a single message. The distance represents the maximum range for transmission between two devices. All data is derived from protocol specifications operating under a 3 MHz clock frequency.
\begin{equation}
\label{for:2}
    EBR=\frac{B_{data}}{T}
\end{equation}

Table~\ref{tab:compa} shows that LinkBo outperforms the other protocols in bit rate and throughput. With two devices, it delivers HP messages in 50.4 µs and incurs a maximum delay of 223.8 µs for LP 7-byte messages. The highest EBR occurs with LP messages, and the lowest with HP. While 1-Wire (ADI) supports long distances (up to 50 m), it requires longer resets and bit slots. UNI/O reduces reset overhead but uses 8-bit synchronization and dual ACKs per byte, increasing latency and lowering EBR. In contrast, LinkBo uses only 2 sync bits and shorter bit slots, reducing message size and latency.

\begin{table}
    \caption{Comparison between 1-wire (ADI), UNI/O and LinkBo.}
    \begin{center}
    \begin{tabular}{c|c|c|c}
    \hline
    Protocol &1-wire (ADI) & UNI/O &\textbf{LinkBo} \\
    \hline
    Type&Asynchronous&Asynchronous&Asynchronous \\
    Duplex&Half &Half & Half\\
    Bit rate (kbps)& 8.33 - 111 & 10 - 100 & \textbf{294.8 - 297.6}\\
    EBR (kbps)& 5.8 - 77.2 & 7.97 - 79.7 & \textbf{158.72 - 252.5}\\
    Latency (µs)& 1520 - 4880 & 810 - 1410 & \textbf{50.4 - 223.8}\\
    \hline
    Interrupt& No & No & \textbf{Yes}\\
    Multi-byte& No & Yes & \textbf{Yes}\\
    CRC&Yes, 8 bits& No &\textbf{Yes, 4 bits} \\
    Acknowledge& No & Yes & \textbf{Yes}\\
    Distance (m)&20-100& N/A & \textbf{5.6-15}\\
    \hline
    \end{tabular}
    \label{tab:compa} 
    \end{center}
    \vspace{-0.5cm} 
\end{table}

\section{Conclusion}
\label{sec:concl}
In this paper, we presented LinkBo, a novel single-wire, low-latency, robust and high-throughput protocol for variable-distance chip-to-chip communication, include its protocol definition and its hardware architecture.
LinkBo can achieve a transmission with different priority and error detection method, using Manchester encoding. High priority and interrupt can reduce minimum delay of transmission.
We designed a novel hardware architecture for LinkBo protocol and implemented on a hardware demonstrator comprising two FPGA boards. 
Our protocol achieves a low latency of just 50.4 µs, which is lower by at least 20X and 6.3X compared to the SOTA 1-wire (ADI) and UNI/O protocol, respectively.
Furthermore, we investigated robustness by changing the wire length and bit rate on FPGA performance, demonstrating that our hardware facilitates communication over distances up to 15~meters at around 300~kbps, and 11~cm at 7.5~Mbps. This capability ensures robustness suitable for different distance from chip-to-chip communication to standard PCB-level communication.

\bibliographystyle{ieeetr}
\bibliography{rf1}

\end{document}